\documentclass[12pt]{article}

\usepackage{latexsym,amsmath,amssymb,amsthm,amsfonts,graphicx,natbib,breqn,epsfig,bm,authblk,url,thmtools}

\usepackage{float} 
\usepackage{booktabs} 
\usepackage{graphicx} 
\usepackage[margin=1cm]{caption} 
\usepackage{mathtools}
\usepackage{enumerate}
\usepackage{bbm}
\usepackage{tikz}
\usepackage[colorinlistoftodos]{todonotes}
\usetikzlibrary{shapes.multipart}
\usetikzlibrary{arrows, positioning}

\floatstyle{plain}
\newfloat{Algorithm}{thp}{lop}
\floatname{Algorithm}{Algorithm}

\declaretheoremstyle[
notefont=\bfseries, notebraces={}{},
bodyfont=\normalfont,
postheadspace=0.5em,
numbered=yes,
]{mystyle}

\theoremstyle{definition}

\begin{document}

\title{Detecting conflicting summary statistics in likelihood-free inference}
\date{\empty}

\author[1]{Yinan Mao}
\author[2]{Xueou Wang\thanks{Corresponding author:  a0095911@u.nus.edu}}
\author[1,3]{David J. Nott}
\author[4]{Michael Evans}
\affil[1]{Department of Statistics and Applied Probability, National University of Singapore, Singapore 117546}
\affil[2]{Institute of Data Science, National University of Singapore, 
Singapore 117602}
\affil[3]{Operations Research and Analytics Cluster, National University of Singapore, Singapore 119077}
\affil[4]{Department of Statistics, University of Toronto, Toronto, ON M5S 3G3, Canada}

\maketitle

\vspace{-0.3in}

\begin{abstract}
Bayesian likelihood-free methods implement Bayesian inference using simulation of data from the model 
to substitute for intractable likelihood evaluations.  
Most likelihood-free inference methods replace the full data set with a summary statistic before
performing Bayesian inference, and the choice of this statistic is often difficult.  
The summary statistic should be low-dimensional for computational reasons, 
while retaining as much information as possible about the parameter.  Using a recent
idea from the interpretable machine learning literature, we develop some regression-based diagnostic methods
which are useful for detecting when different parts of a summary statistic vector contain conflicting information 
about the model parameters. Conflicts of this kind complicate summary statistic choice, and detecting them can be insightful about
model deficiencies and guide model improvement.  The diagnostic methods developed are based on regression approaches to likelihood-free inference, in which the regression model estimates the posterior
density using summary statistics as features.  Deletion and imputation of part of the summary statistic vector
within the regression model can remove conflicts and approximate posterior distributions for
summary statistic subsets.  A larger than expected change in the estimated posterior density
following deletion and imputation can indicate a conflict in which inferences of interest are affected.  
The usefulness of the new methods is demonstrated in a number of real examples.  

\smallskip
\noindent \textbf{Keywords.}  Approximate Bayesian computation; Bayesian model criticism; influence measures;  likelihood-free inference; model misspecification.

\end{abstract}

\section{Introduction}\label{sec:Intro}

Often scientific knowledge relevant to statistical model development comes in the form of a possible generative process
for the data.  Computation of the likelihood is sometimes intractable for models specified in this way, and then
likelihood-free inference methods
can be used if simulation of data from the model
is easily done.  A first step in most likelihood-free inference
algorithms is to reduce the full data set to a low-dimensional summary statistic, which 
is used to define a distance for comparing observed and simulated data sets.    
The choice of this summary statistic must be done carefully, balancing computational and 
statistical considerations \citep{blum+nps13}.  
Once summary statistics are chosen, Bayesian likelihood-free inference can proceed using a variety
of techniques, such as approximate Bayesian computation (ABC) \citep{marin+prr12,sisson+fb18} synthetic likelihood
\citep{wood10,price+dln16} or regression based approaches \citep{beaumont+zb02,blum+f10,fan+ns13,raynal+mprre18}.

Sometimes different components of the summary statistic vector bring conflicting information
about the model parameter. This can arise due to model deficiencies that we wish to be aware of, and can complicate summary statistic choice.   Furthermore, it can lead to computational difficulties when observed summary statistics 
are hard to match under the assumed model.  We develop some diagnostics for detecting when
this problem occurs, based on a recent idea in the interpretable machine learning literature.  The suggested diagnostics
are used in conjunction with regression-based approaches to likelihood-free inference.  

The following example of inference on a Poisson mean, which is discussed in \cite{sisson+fb18intro}, 
illustrates the motivation for our work.    Here the likelihood is tractable, but the example is useful for pedagogical purposes.
Suppose data $y=(y_1,\dots, y_5)^\top=(0,0,0,0,5)^\top$ are observed, and that we model $y$ as a Poisson random
sample with mean $\eta$.  For Bayesian inference we use a $\text{Gamma}(1,1)$ prior density for $\eta$.  The gamma prior is conjugate, and the posterior density is 
$\text{Gamma}(1+5\bar{y},6)$, where $\bar{y}$ is the sample mean of
$y$.  Letting $s^2$ denote the sample variance of $y$, 
both $\bar{y}$ and $s^2$ are point estimates of $\eta$, since for the Poisson model the mean and 
variance are equal to $\eta$;  furthermore, $\bar{y}$ is a minimal sufficient statistic.  

Consider a standard rejection ABC method for this problem, which involves generating $\eta$ values from the prior, generating summary statistics under the model conditional on the generated $\eta$ values, and then keeping the $\eta$ values which lead to simulated summary statistics closest to those observed.    See \cite{sisson+fb18intro} for further discussion
of basic ABC methods.
Consider first the summary statistic $\bar{y}$, and apply rejection ABC with $100,000$ prior samples and 
retaining $500$ of these in the ABC rejection step.  
This gives a very close approximation to the true posterior (Figure 1). 
\begin{figure}[h]
\centerline{\includegraphics[width=90mm]{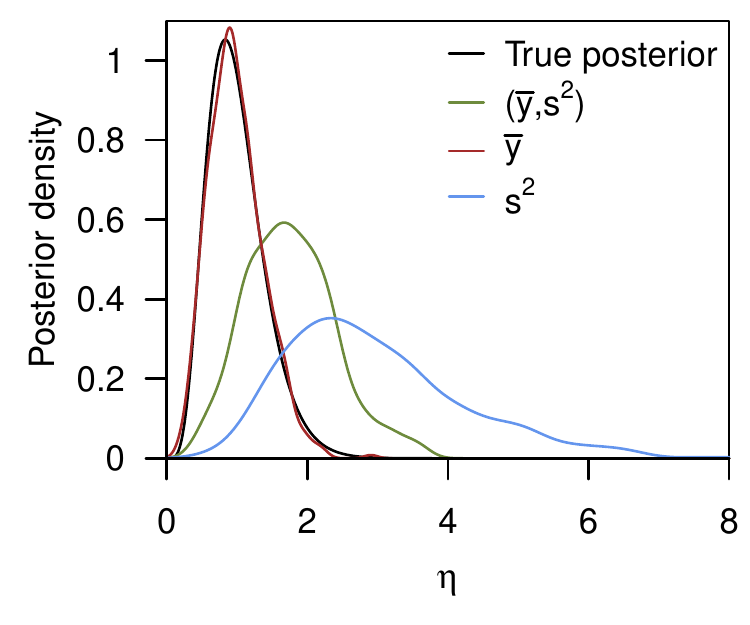}}
\caption{\label{summstatex}Estimated posterior densities via rejection ABC with 100,000 model simulations and retaining $500$
samples in Poisson means example with different summary statistic choices.}
\end{figure}
The computations were implemented using the default implementation of rejection ABC in
the R package {\tt abc} \citep{csillery+fb12}.
Reduction of the full data to $\bar{y}$ involves no loss of information, as $\bar{y}$ is sufficient.
  
Since we observe values of $\bar{y}=1$ and $s^2=5$, these summaries bring conflicting information about
$\eta$, since they are both point estimates of $\eta$ and take quite different values.   If we consider the summary statistic $(\bar{y},s^2)$, and apply the same rejection ABC algorithm as before with $100,000$ prior samples and $500$ accepted samples, 
the estimated posterior
density changes appreciably -- see Figure 1.  Although $(\bar{y},s^2)$ is still sufficient, and the corresponding posterior 
is the same as the posterior given $\bar{y}$, the ABC computational algorithm performs poorly because the observed 
summary statistic is hard to match with a reasonable fixed computational budget.  
Simultaneously observing $\bar{y}=1$ and $s^2=5$ is 
unusual under the assumed model for any value of the model parameter.  
This results in accepting parameter samples having simulated summary statistic values far away from
the observed  value of $(\bar{y},s^2)$.
Also shown in Figure 1 is the posterior conditional on 
$s^2$, which is even further from the true posterior than the posterior conditional on $(\bar{y},s^2)$.

The conflicting information in the summary statistics $\bar{y}$ and $s^2$ may indicate a model failure, motivating
a more flexible model than the Poisson, such as negative binomial, in which the mean and variance are not
required to be equal.  \cite{sisson+fb18intro} considered this example partly as a demonstration of the difficulties which arise in 
greedy algorithms for summary statistic choice.  An analyst following a policy of adding one summary statistic at a time, and
stopping when the posterior density no longer changes appreciably, would incorrectly judge $(\bar{y},s^2)$ to be a better
summary statistic than $\bar{y}$.   
Our example illustrates that if different parts of a summary statistic vector are conflicting, then detecting this
can be insightful about model failures and can suggest model improvements.  This is the motivation
for the conflict detection diagnostics developed here.  

The structure of the paper is as follows.  In the next Section we
describe regression approaches to likelihood-free inference.  Section 3 then 
discusses a recent suggestion in the literature on interpretable
machine learning, which has the goal of explaining how a flexible classifier makes the class prediction it does for a certain observation.
This method is based on deletion of part of a feature vector, followed by imputation of what was deleted and assessment of 
how evidence for the observed class changes following the deletion and imputation process.  
In regression approaches to computing likelihood-free inferences, where regression models are considered
with simulated parameters as the response values and simulated summary statistics as features,  
we show that a similar approach can be used to estimate the changes in a posterior
distribution under sequential updating for summary statistic subsets.   If the change in the posterior
density is larger than expected when a subset of summary statistics is deleted, this may
indicate that different parts of the summary statistic vector carry conflicting information about the parameter.

In Section 4, our diagnostics are formalized 
and placed in the broader framework of Bayesian model checking.   We make connections with the literature
on prior-data conflict, suggest a 
method of calibration, and also review related literature on model checking for likelihood-free methods.   
Section 5 considers three examples.  First we revisit the pedagogical example above and show how 
 our diagnostics apply in that simple case, 
before considering two further real examples.  The first real example considers
data on stereological extremes, where the model is not flexible enough to fit well in both the
lower and upper tails.  The second example concerns an ecological time series model.  An  
analysis of these data using summary statistics derived from an auxiliary threshold autoregressive model
is first considered, where
the summary statistics can be thought of as capturing the different 
dynamic behaviour of the time series at high and low levels.  
Conflicts between summary statistic subsets for different regimes 
are insightful for understanding deficiencies in the original model.  An analysis of these data
without summary statistics is also considered, 
where the regression model is based on a convolutional neural network, and the feature
vector consists of the whole time series.  Here we explore temporally localized parts of the series
not fitted well by the model.
Section 6 gives some concluding discussion.

\section{Regression approaches to likelihood-free inference}

Consider a Bayesian inference problem with data $d$ and a parametric model for $d$ with
density $p(d|\eta)$, with parameter $\eta$.  The prior density for $\eta$ is denoted $p(\eta)$, 
with corresponding posterior density 
$p(\eta|d_{\text{obs}})\propto p(\eta,d_{\text{obs}})=p(\eta)p(d_{\text{obs}}|\eta)$, where $d_{\text{obs}}$
denotes the observed value  of $d$.  Suppose that computation of the likelihood $p(d_{\text{obs}}|\eta)$ is intractable, 
so that likelihood-free inference approaches are needed to approximate the posterior distribution.  
In this case, we can consider summary statistics $S=S(d)$, with $S_{\text{obs}}=S(d_{\text{obs}})$, and 
writing $p(S|\eta)$ for the sampling density of $S$, we 
use likelihood-free inference methods to approximate 
$p(\eta|S_{\text{obs}})\propto p(\eta)p(S_{\text{obs}}|\eta)$.  

The posterior density $p(\eta|S_{\text{obs}})$ is the conditional
density of $\eta$ given $S=S_{\text{obs}}$ in the joint Bayesian model for $(\eta,S)$ 
specified by $p(\eta,S)=p(\eta)p(S|\eta)$.  Since regression is conditional density estimation, we can fit
regression models to samples $(\eta^{(i)},S^{(i)})$, $i=1,\dots, N$ from $p(\eta,S)$ in order 
to estimate $p(\eta|S)$ for any $S$.  
The responses are $\eta^{(i)}$, and the features are $S^{(i)}$.  
The regression predictive density given $S=S_{\text{obs}}$, denoted $\widetilde{p}(\eta|S_{\text{obs}})$, is an estimate
of $p(\eta|S_{\text{obs}})$.  

There are many flexible regression approaches that
have been applied to computation of likelihood-free inferences, such as quantile regression forests 
\citep{meinshausen06,raynal+mprre18}, copula methods \citep{li+nfs15,fan+ns13,klein+ns20}, 
neural approaches of various kinds \citep{papamakarios+m16,papamakarios+sm19} and orthogonal series methods
\citep{izbicki+l17,izbicki+lp19} among others.  
In our later applications to the detection of conflicting information in summary statistics we will be concerned with the effects 
of conflict on inference for
scalar functions of the parameter, and so only univariate regression models are needed for estimating 
one-dimensional marginal posterior distributions.  
To save notation we will use the notation $\eta$ both for the model parameter as well as a one-dimensional function
of interest of it, where the meaning intended will be clear from the context.

In checking for summary statistic conflicts, we will 
partition $S$ as $S=(S_A^\top,S_B^\top)^\top$ and we ask if the information about $\eta$ in
$S_A$ conflicts with that in $S_B$.  We write $S_{\text{obs}}=(S_{A,\text{obs}}^\top,S_{B,\text{obs}}^\top)^\top$.  
Our conflict diagnostic will compare an estimate of $p(\eta|S_{\text{obs}})$
to an estimate of $p(\eta|S_{A,\text{obs}})$.  If the change in the estimated posterior distribution is large, 
this may indicate that the summary statistics carrying conflicting information about the parameter $\eta$.  
We must define what large means, and this is an issue of calibration of the diagnostics.

To overcome the computational difficulties of estimating the posterior
distributions, we will first consider a likelihood-free regression method with features $S$ to compute $p(\eta|S_{\text{obs}})$.  
Then based on this same regression, 
the method of \cite{zintgraf+caw17} described in Section 3 will be used to 
approximate $p(\eta|S_{A,\text{obs}})$ by deletion and multiple imputation of $S_B$.  
This can be done without refitting
the regression.  This last point is important, because sometimes fitting very flexible regression models
can be computationally burdensome.  For example, in the application of Section 5.2, we consider a training
sample of $500,000$ observations, where we fit a quantile regression forest in which algorithm fitting parameters
are tuned by cross-validation.  Applying the method of \cite{zintgraf+caw17} for approximating the posterior
density given $S_{A,\text{obs}}$, using the fitted regression with features $S$,
is much less computationally demanding than refitting the regression with features $S_A$
to approximate the subset posterior density directly.  Since we may wish to consider many different
choices of $S_A$ and $S_B$ for diagnostic purposes, this is another reason to avoid the approach
of refitting the regression model with different choices of the features.

Although the main idea of our method is to compare estimates of $p(\eta|S_{\text{obs}})$ and $p(\eta|S_{A,\text{obs}})$, 
making this precise will involve the development of some further background.  We describe the
method of \cite{zintgraf+caw17} next, and then in Section 4 we formalize our diagnostics
and consider connections to the existing literature on Bayesian model checking.

\section{Interpretable machine learning using imputation}\label{interpretableml}

We describe a method developed in \cite{zintgraf+caw17} for visualizing classifications made by a 
deep neural network.  Their approach 
builds on earlier work of \cite{robnik+k08}.  
The method deletes part of a feature vector, and then replaces 
the deleted part with an imputation based on the other features.  The change in the 
observed class probability is then examined to measure of how important
the deleted features were to the classification made.  To avoid repetition,
the discussion below will be concerned with a general regression model where the response variable can be continuous
or discrete, although \cite{zintgraf+caw17} deals only with the case where the response is a discrete
class label.  It is the case of regression models with a continuous response that is relevant to our
discussion of likelihood-free inference.

Suppose that $(y_i,x_i)$, $i=1,\dots, n$ are response and feature vector pairs in some training set of $n$ observations.  
The feature vectors are $p$-dimensional, written as $x_i=(x_{i1},\dots, x_{ip})^\top$.  We assume the responses
are scalar valued, although extension to multivariate responses is possible.  Write $Y=(y_1,\dots, y_n)^\top$ and 
$X=[x_1^\top, \dots, x_n^\top]^\top$ so that $Y$ is an $n$-vector and $X$ is an $n\times p$ matrix.  
We consider Bayesian inference in this regression model with a prior density $p(\theta)$ for $\theta$, and
likelihood $p(Y|\theta,X)$.   Note the different notation here to Section 2, where we considered a likelihood-free inference
problem with data $d$ and parameter $\eta$.  Here $Y$ and $X$ are the responses and features in
the regression, and in likelihood-free regression the features $X$ will be simulated summary statistics, 
whereas the responses $y$ are simulated values of the parameters $\eta$, hence
the different notation.   More precisely, in the likelihood-free problems of interest to us, and in the notation of Section 2, 
$y_i=\eta^{(i)}$ and $x_i=S^{(i)}$, $i=1,\dots, n$.  

The posterior density of $\theta$ in the regression model is
\begin{align*}
  p(\theta|Y,X) & \propto p(\theta)p(Y|\theta,X).
\end{align*}
For some feature vector $x^*$, and corresponding value of interest for the response $y^*$, 
we can compute the predictive density of $y^*$ given $x^*,Y,X$ as
\begin{align*}
  p(y^*|x^*,Y,X) & = \int p(y^*|\theta,x^*)p(\theta|Y,X) d\theta,
\end{align*}
where $p(y^*|\theta,x^*)$ is the likelihood term for $y^*$ and it is assumed that $(y^*,x^*)$ is conditionally independent
of $(Y,X)$ given $\theta$ and $x^*$.  

Suppose that some subector of $x^*$ is not observed, say the $i$th component $x_i^*$.  Assume
that the features $x^*,x_1,\dots, x_n$ are modelled as independent and identically distributed from a density
$p(x|\lambda)$, where the parameters $\lambda$ are disjoint from $\theta$, and that $\lambda$ and $\theta$ 
are independent in the prior.  We write $x^*_{-i}$ for $x^*$ with $x_i^*$ deleted.  Then
under the above assumptions,
\begin{align}
  p(y^*|x^*_{-i},X,Y) 
   & = \int p(y^*|x^*,Y,X) p(x_i^*|x_{-i}^*,X) dx_i^*,  \label{imputation}
\end{align}
where
\begin{align*}
  p(x_i^*|x_{-i}^*,X) & = \int p(x_i^*|x_{-i}^*,\lambda)p(\lambda|X) d\lambda.
\end{align*}
Equation (\ref{imputation}) says that to get $p(y^*|x_{-i}^*,Y,X)$ we can average 
$p(y^*|x^*,y,X)$ over the distribution of $x_i^*$ where $x_i^*$ is imputed from $x_{-i}^*$. 
In the discussion above only deletion of the $i$th component $x_i^*$ of $x^*$
has been considered, but 
the generalization to removing arbitrary subsets of $x^*$ is immediate.
In the likelihood-free problems discussed later, the expression (\ref{imputation}) will
give us a way of estimating the posterior distribution of $\eta$ based on summary statistic
subsets using a regression fitted using the full set of summary statistics and a convenient
imputation model.  We will partition the summary statistic $S$ as $S=(S_A^\top, S_B^\top)^\top$, 
and then with $y^*=\eta$ and $x^*=S_{\text{obs}}$ where $S_{\text{obs}}=(S_{\text{A,obs}}^\top ,S_{\text{B,obs}}^\top)^\top$ is the observed value of $S$, 
the formula (\ref{imputation}) will give us a way of estimating the posterior distributon $p(\eta|S_{A,\text{obs}})$
based on the regression predictive distribution using features $S$.
 
In practice we might approximate $p(x_i^*|x_{-i}^*,X)$ as $p(x_i^*|x_{-i}^*,\hat{\lambda})$ for some point estimate
$\hat{\lambda}=\hat{\lambda}(X)$ of $\lambda$.  For example, we might assume a multivariate normal model for $x_1,\dots, x_n$ (in which case
$\lambda$ is the mean and covariance matrix of the normal model) and plug in the sample mean and covariance matrix
as the point estimate.  This is the approach considered in \cite{zintgraf+caw17}.  \cite{robnik+k08} 
assumed the features to be independent in their model.  \cite{zintgraf+caw17} note that more
sophisticated imputation methods can be used.  
We can also approximate $p(y^*|x^*,Y,X)$ by $p(y^*|x^*,\hat{\theta})$ for some point estimate
$\hat{\theta}$ of $\theta$.  

In the case of a classification problem where the response is a class label, we can follow \cite{zintgraf+caw17}
and \cite{robnik+k08} and measure how much class probabilities have changed upon deletion and imputation
of $x_i^*$ using an evidence measure, called weight of evidence:  \\
\begin{align}
  \text{WE}(y^*|x^*) & = \log_2 \left\{ \frac{p(y^*|x^*,Y,X)}{1-p(y^*|x^*,Y,X)}\right\}-
    \log_2\left\{ \frac{p(y^*|x_{-i}^*,Y,X)}{1-p(y^*|x_{-i}^*,Y,X)}\right\}. \label{wofev}
\end{align}
$\text{WE}(y^*|x^*)$ is positive (negative) if observing $x_i^*$ rather than imputing it from $x_{-i}^*$ makes the class $y^*$ more
(less) probable.  \cite{zintgraf+caw17} generally consider the observed class for $y^*$.  Measures of evidence
suitable for the regression context are considered further in the next section.  

\cite{zintgraf+caw17} consider problems in image classification where the feature vector $x^*$ is a vector of intensities
of pixels in an image.  Because of the smoothness of images,  if only one pixel is deleted, then it can be imputed very
accurately from its neighbours.  If this is the case, $\text{WE}(y^*|x^*)$ is not useful for measuring the importance
of a pixel to the classification.  \cite{zintgraf+caw17} suggest to measure pixel importance by deleting all image patches of a certain 
size, and then average weight of evidence measures for patches that contain a certain pixel.  They plot images of the averaged weight of evidence measure as a way of visualizing the importance of different parts of an image for a classification made
by the network.  They also consider deletion and imputation for latent quantities in deep neural network models.  

An appropriate imputation method will vary according to the problem at hand, 
and there is no single method that will be appropriate for all problems.  In our later examples 
for computing likelihood-free inferences we consider three different imputation methods.
The first uses linear regression in a situation where the features are approximately linearly related, and 
we use the implementation in the \texttt{mice} package in \texttt{R} \citep{vanbuuren+g11}.  
The second is based on random
forests, as implemented in the \texttt{missRanger} package \citep{mayer19}.  
Our final example considers time series data, 
and in the application where we consider the raw series as the feature vector, we use the methods in the 
\texttt{imputeTS} package \citep{steffen+b17}.  Further details are given in Section 5.

\section{The diagnostic method} \label{regressionlf}

In the notation of Section 2, the method of \cite{zintgraf+caw17} will allow us to 
estimate the posterior density $p(\eta|S_{A,\text{obs}})$, from a regression using the full set of 
summary statistics $S$ as the features.  This is described in more detail in 
Section 4.2 below.  We can then compare estimates of $p(\eta|S_{A,\text{obs}})$ and
$p(\eta|S_{\text{obs}})$.   If the difference between these two posterior distributions is large, 
then it indicates that different values of the parameter are providing a good fit to the data
in the likelihood terms $p(S_{A,\text{obs}}|\eta)$ and $p(S_{B,\text{obs}}|S_{A,\text{obs}},\eta)$, 
indicating a possible model failure.  The weight of evidence measure of \cite{zintgraf+caw17} 
doesn't apply directly as
a way of numerically describing the difference in the posterior distributions 
$p(\eta|S_{A,\text{obs}})$ and $p(\eta|S_{\text{obs}})$ in the likelihood-free
application.  Here we use measures of discrepancy considered in the prior-data conflict checking
literature \citep{nott+wee20} based on the concept of relative belief \citep{evans15}  instead of weight of evidence.  
We place our diagnostics within the framework of Bayesian model checking, and this suggests a method for calibrating the diagnostics.  The discussion of prior-data conflict checking is relevant here, because the comparison
of $p(\eta|S_{\text{obs}})$ to $p(\eta|S_{A,\text{obs}})$ can be thought of as a prior-to-posterior comparison
in the situation where $S_A$ has already been observed, and then $p(\eta|S_{A,\text{obs}})$ is the prior for
further updating by the likelilhood term $p(S_{B,\text{obs}}|S_{A,\text{obs}},\eta)$.  A more detailed
explanation of this is given below.  

\subsection{Prior-data conflict checking}

Prior-data conflicts occur if there are values of the model parameter that receive likelihood support, but the prior
does not put any weight on them \citep{evans+m06,presanis+osd13}.  In other words, prior-data conflict occurs when 
the prior puts all its mass out in the tails of the likelihood.  
\cite{nott+wee20} consider some methods for checking for prior-data conflict
based on prior-to-posterior divergences.  
In the present setting of a posterior density $p(\eta|S)$ for summary statistics $S$ and with prior $p(\eta)$, the
checks of \cite{nott+wee20} 
would check for conflict between $p(\eta)$ and $p(S|\eta)$ using a prior-to-posterior $\alpha$-divergence \citep{renyi61} taking the form
\begin{align}
  R_\alpha(S) & = \frac{1}{\alpha-1} \log \int \left\{ \frac{p(\eta|S)}{p(\eta)}\right\}^{\alpha-1} p(\eta|S) \,d\eta, \label{renyi-div}
\end{align}
where $\alpha>0$.  These divergences are related to relative belief functions \citep{evans15} measuring evidence
in terms of the prior-to-posterior change.  
Consider a function $\psi=T(\eta)$ of $\eta$.  We define the relative belief function for $\psi$ as
\begin{align}
  \text{RB}(\psi|d) & = \frac{p(\psi|d)}{p(\psi)}=\frac{p(d|\psi)}{p(d)}, \label{relative-belief}
\end{align}
where $p(\psi|d)$ is the posterior density for $\psi$ given $d$, $p(\psi)$ is the prior density, $p(d|\psi)$ is the marginal
likelihood given $\psi$, and $p(d)$ is the prior predictive density at $d$.  For a given value of 
$\psi$, if $\text{RB}(\psi|d)>1$, this means there is evidence in favour of $\psi$, whereas if $\text{RB}(\psi|d)<1$ there is evidence
against.  See \cite{evans15}
for a deeper discussion of reasons why relative belief is an attractive measure of evidence, and for a systematic approach
to inference based on it.  

As $\alpha\rightarrow 1$, (\ref{renyi-div}) gives the prior-to-posterior Kullback-Leibler divergence (which
is the posterior expectation of $\log \text{RB}(\eta|S)$), and letting $\alpha\rightarrow\infty$ 
gives the maximum value of the log relative belief, $\log \text{RB}(\eta|S)=\log p(\eta|S)/p(\eta)$.  So 
the statistic (\ref{renyi-div}) is a measure of the overall size of a relative belief function, and 
a measure of how much our beliefs have changed from prior to posterior.  \cite{nott+wee20}
suggest to compare the value of $R_\alpha(S_{\text{obs}})$ to the distribution of $R_\alpha(S)$, $S\sim p(S)$
where $p(S)$ is the prior predictive density for $S$, 
\begin{align*}
 p(S) & = \int p(S|\eta)p(\eta)\,d\eta,
\end{align*}
by using the tail probability
\begin{align}
  p_S & = P(R_\alpha(S)\geq R_\alpha(S_{\text{obs}})). \label{calibration}
\end{align}
The tail probability (\ref{calibration}) is the probability that the overall size
of the relative belief function RB$(\eta|S_{\text{obs}})$ is large compared to what is expected for data following
the prior predictive density $p(S)$.  So a small value for the tail probability (\ref{calibration}) suggest the prior and likelihood
contain conflicting information, as the change from prior to posterior is larger than expected.  
See \cite{evans+m06} and \cite{nott+wee20} for further discussion of prior-data conflict
checking, and why the check in \cite{nott+wee20} satisfies certain logical requirements for such a check.

\subsection{Definition of the diagnostics}

In the present work, we have a summary statistic $S$ partitioned as 
$S=(S_A^\top,S_B^\top)^\top$, and it is conflict between $S_A$ and $S_B$ that interests us.  
This is related to prior-data conflict checking in the following way.  Suppose that we have the posterior
density $p(\eta|S_A)$.  
We can regard $p(\eta|S_A)$ as a prior density to be updated by the likelihood term
$p(S_B|S_A,\eta)$ to get the posterior density $p(\eta|S)$.  That is, 
$p(\eta|S) \propto p(\eta|S_A) p(S_B|S_A,\eta)$.
The density $p(\eta|S_A)$ usefully reflects information
in $S_A$ about $\eta$, if we have checked both the model for $p(S_A|\eta)$ and 
for prior-data conflict between $p(\eta)$ and $p(S_A|\eta)$ in conventional ways.  
If updating $p(\eta|S_A)$ by the information in $S_B$ gives a posterior density 
$p(\eta|S)$ that is surprisingly far from $p(\eta|S_A)$, this
indicates that $S_A$ and $S_B$ bring conflicting information about $\eta$.  
The appropriate reference distribution
for the check is 
\begin{align*}
  p(S_B|S_A) & = \int p(S_B|S_A,\eta)p(\eta|S_A) d\eta, 
\end{align*}
since we are conditioning on $S_A$ in the prior that reflects knowledge of $S_A$ before
updating for $S_B$.

Suppose that the regression model used for the likelihood-free inference gives an approximate posterior density
$\widetilde{p}(\eta|S)$ for summary statistic value $S$. The change 
in the approximate posterior when deleting $S_{B,\text{obs}}$ from $S_{\text{obs}}$ can be examined 
using a mulitple imputation
procedure, using the method of \cite{zintgraf+caw17}.  
We produce $M$ imputations $S_B(i)$, $i=1,\dots, M$ of $S_B$ and writing
$S(i)=(S_{A,\text{obs}}^\top, S_B(i)^\top)^\top$ and following (\ref{imputation})
we approximate $p(\eta|S_{A,\text{obs}})$ by
\begin{align*}
 \widetilde{p}(\eta|S_{A,\text{obs}}) & = \frac{1}{M}\sum_{i=1}^M \widetilde{p}(\eta|S(i)).
\end{align*}
The change in the approximate posterior density from
$\widetilde{p}(\eta|S_{A,\text{obs}})$ to $\widetilde{p}(\eta|S_{\text{obs}})$ can be
summarized by the maximum log relative belief
\begin{align}
 R_\infty(S_{B,\text{obs}}|S_{A,\text{obs}}) & = \sup_\eta \log \frac{\widetilde{p}(\eta|S_{\text{obs}})}{\widetilde{p}(\eta|S_{A,\text{obs}})}.  \label{logrbstat}
 \end{align}
 Our discussion of the prior-data conflict checks of \cite{nott+wee20} suggests that an appropriate reference distribution
 for calibrating a Bayesian model check using (\ref{logrbstat}) would compute the tail probability
 \begin{align}
   p & = P(R_\infty(S_B|S_{A,\text{obs}}) \geq R_\infty (S_{B,\text{obs}}|S_{A,\text{obs}})), \label{pvalue}
\end{align}
where $S_B\sim p(S_B|S_{A,\text{obs}})$.  The imputations in the multiple imputation procedure approximate
draws from $p(S_B|S_{A,\text{obs}})$, and 
we suggest drawing a fresh set of imputations $S_B^*(i)$, $i=1,\dots, M^*$, separate
from those used in computing $\widetilde{p}(\eta|S_{A,\text{obs}})$, and to approximate (\ref{pvalue}) by 
\begin{align}
  \widetilde{p} & = \frac{1}{M^*} \sum_{i=1}^{M^*} I(R_\infty(S_B^*(i)|S_{A,\text{obs}})\geq 
    R_\infty(S_{B,\text{obs}}|S_{A,\text{obs}})),  \label{p-value}
\end{align}
where $I(A)$ is the indicator function which is $1$ when event $A$ occurs and zero otherwise.
Our suggested checks use the statistic (\ref{logrbstat}), calibrated using the estimated tail probability 
(\ref{p-value}).  

\subsection{The logic of Bayesian model checking}

\cite{evans+m06} discuss a strategy for model checking based on a decomposition of the joint Bayesian
model $p(\eta,d)=p(\eta)p(d|\eta)$, generalizing an earlier approach due to \cite{box80}.  
Different terms in the decomposition should be used for different purposes such as inference 
and checking of different model components.  We do not discuss their approach in detail, but
an important principle is that for checking model components we should check the sampling model first, and only afterwards
check for prior-data conflict.  The reason is that if the model for the data is inadequate, 
sound inferences cannot result no matter what the prior is.  

In our analysis, the posterior distribution for $\eta$ is
\begin{align}
  p(\eta|S_{\text{obs}}) & \propto p(\eta)p(S_{\text{obs}}|\eta). \label{full-update}
\end{align}
Our diagnostics consider sequential Bayesian updating in two stages, 
\begin{align}
  p(\eta|S_{A,\text{obs}}) & \propto p(\eta) p(S_{A,\text{obs}}|\eta), \label{stage-one-update}
\end{align}
and 
\begin{align}
  p(\eta|S_{\text{obs}}) & \propto p(\eta|S_{A,\text{obs}})p(S_{B,\text{obs}}|S_{A,\text{obs}},\eta). \label{stage-two-update}
\end{align}
In (\ref{stage-one-update}), we can check the sampling model $p(S_A|\eta)$ and then check for conflict
between $p(\eta)$ and $p(S_A|\eta)$, and if these checks are passed, 
then $p(\eta|S_A)$ usefully reflects information about $\eta$ in 
$p(S_A|\eta)$.  We could then proceed to check the components in (\ref{stage-two-update}). 
Checking $p(S_A|\eta)$ in (\ref{stage-one-update}) and then $p(S_B|S_A,\eta)$ in (\ref{stage-two-update}) is not
the same as checking $p(S|\eta)$ in (\ref{full-update}).  The reason is that there can be values of $\eta$ providing
a good fit to the data for $p(S_A|\eta)$ and values of $\eta$ providing a good fit to the data in $p(S_B|S_A,\eta)$, 
but these need not be the same parameter values.  

Formally, we could imagine an expansion of the original model from
\begin{align}
  p(S|\eta) & = p(S_A|\eta)p(S_B|S_A,\eta), \label{model-expansion}
\end{align}
to
\begin{align}
  p(S|\eta_1,\eta_2) & = p(S_A|\eta_1)p(S_B|S_A,\eta_2), 
\end{align}
so that the original parameter $\eta$ is now allowed to vary between 
the two likelihood terms, with the original model corresponding to $\eta_1=\eta_2$.  
If a good fit to the data can only be obtained 
with $\eta_1\neq \eta_2$, one would expect that our diagnostics would detect this, as the likelihood
terms are peaked in different regions of the parameter space, and this will become evident in the sequential
updating of the posterior for the different summary statistic subsets. 

The inherent asymmetry in $S_A$ and $S_B$ in the decomposition (\ref{model-expansion})  
has consequences for interpretation of the diagnostics.  To explain this, 
consider once more the toy Poisson model in the introduction.  Consider first $S_A=\bar{y}$ and 
$S_B=s^2$.  Since $\bar{y}$ is sufficient, we have $p(\eta|S_A)=p(\eta|S)$, and our diagnostics
would not detect any conflict.  However, conventional model checking of the likelihood
term $p(S_B|S_A,\eta)$ would detect that the model fits poorly in this case.  On the other hand, if
$S_A=s^2$ and $S_B=\bar{y}$, then the term $p(S_B|S_A,\eta)$ depends on $\eta$ since
$s^2$ is non-sufficient, and $p(\eta|S)$ is very different to $p(\eta|S_A)$.  In this case, our diagnostics
do indeed reveal the conflict as we show later in Section 5.1.  
The idea of a conflict between summary statistics is being formalized here in terms of conflict between
the likelihood terms $p(S_A|\eta)$ and $p(S_B|S_A,\eta)$ and this is not symmetric in $S_A$ and $S_B$. 
There is nothing wrong with this, but the interpretation of the diagnostics needs to be properly
understood.

\subsection{Other work on model criticism for likelihood-free inference}

We discuss now related
work on Bayesian model criticism for likelihood-free inference, which is an active area of current research.  
Theoretical examinations of 
ABC methods under model misspecification have been given recently by \cite{ridgway17} and \cite{frazier+rr20}.  
The latter authors contrast the behaviour of rejection ABC and regression adjusted ABC methods \citep{beaumont+zb02,
blum+f10,li+f18} in the case of incompatible summary statistics, where the observed summary statistics
can't be matched by the model for any parameter.  Regression adjustment methods can behave in undesirable ways 
under incompatibility, and the authors suggest a modified regression adjustment for this case.  They develop 
misspecification diagnostics based on algorithm acceptance probabilities, and comparing
rejection and regression-adjusted ABC inferences.  

\cite{ratmann+awr09} considered model criticism based on a reinterpretation of the algorithmic tolerance parameter
in ABC.  They consider a projection of data onto the ABC error, and use a certain predictive density
for this error for model criticism.  \cite{ratmann+prr11} gives a succinct discussion of the fundamentals of the approach and 
computational aspects of implementation.   In synthetic likelihood approaches to likelihood-free inference
\citep{wood10,price+dln16} recent work of \cite{frazier+d19} has considered some model expansions
that are analogous to the method of \cite{ratmann+prr11} for the synthetic likelihood.  Their method
makes Bayesian synthetic likelihood methods less sensitive to assumptions, and the posterior distribution
of the expansion parameters can also be insightful about the nature of any misspecification.
The work of \cite{wilkinson13}, although not focusing on model criticism, is also related to \cite{ratmann+prr11}.
\cite{wilkinson13} shows how the ABC tolerance can be thought of
as relating to an additive ``model error'', and that ABC algorithms give exact
results under this interpretation.    \cite{thomas+pslkc20} discuss Bayesian optimization approaches to learning high-dimensional
posterior distributions in the ABC setting in the context of the coherent loss-based Bayesian inference
framework of \cite{bissiri+hw16}.  Their work
is focused on robustness to assumptions and computation in high-dimensions, rather than on methods for
diagnosing model misspecification.  Recent work of \cite{frazier+dl20} considers a robust ABC approach
related to the model expansion synthetic likelihood method of \cite{frazier+d19}.  This method can also be applied
in conjunction with regression adjustment, and the regression adjustment approach can have better behaviour
in the case of misspecification.  

Also relevant to our work is some of the literature on summary statistic choice.  
In particuclar, \cite{joyce+m08} have considered an approach to this problem which
examines changes in an estimated posterior density upon addition of a new summary statistic.  This is a useful method
for a difficult problem, but their method
is not used as a diagnostic for model misspecification, and in fact can sometimes perform poorly in exactly this situation. 

In a sense, model criticism for Bayesian likelihood-free inference is conceptually no different to Bayesian model criticism with a
tractable likelihood.  However, issues of computation and summary statistic choice sometimes justify the use of different methods.
For general discussions of principles of Bayesian model checking see \cite{gelman+ms96}, \cite{bayarri+c07} and
\cite{evans15}.

\section{Examples}\label{examples}

\subsection{Poisson model}

We return to the toy motivating example given in the introduction on inference for a Poisson mean, where
data $y=(y_1,\dots, y_5)^\top=(0,0,0,0,5)^\top$ are observed.  For these data the sample mean 
is $\bar{y}=1$ and the sample variance is $s^2=5$.  In the Poisson model,
$\bar{y}$ and $s^2$ are both point estimates of the
Poisson mean $\eta$.  The estimates bring conflicting information about the parameter, because
simultaneously observing $\bar{y}=1$ and $s^2=5$ is unlikely for any value of $\eta$.  
Figure 1 in the introduction showed ABC estimates of the posterior density for different summary statistic choices, for the prior
described in Section 1.  Recall that the rejection ABC algorithm
performs poorly for the summary statistic $(\bar{y},s^2)^\top$, because of the difficulty of matching the
observed values, and this may indicate a poorly fitting model.  

\begin{figure}[h]
	\centerline{\includegraphics[width=90mm]{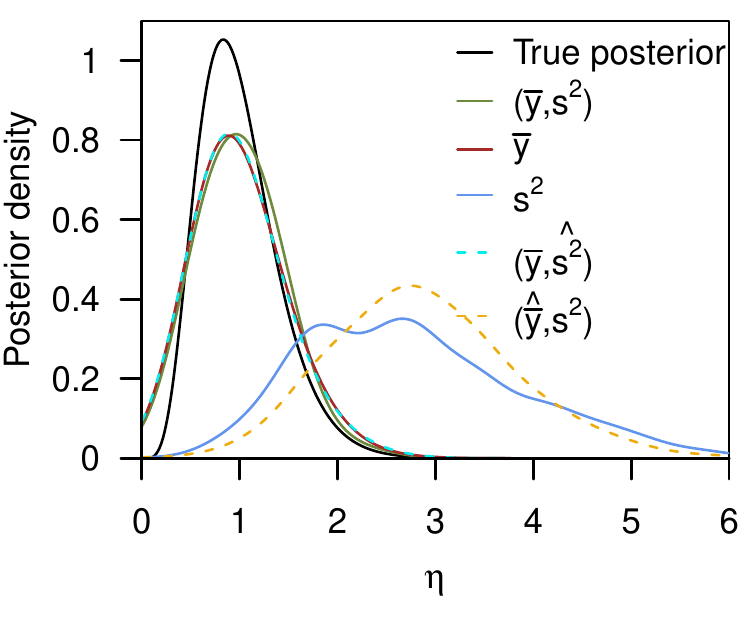}}
	\caption{Estimated posterior density of $\eta$ using ABC random forests with different sets of summary statistics in Example 5.1. }
	\label{fig:eg1_abcrf}
\end{figure}

We apply the diagnostic method of Section 4 for this example, to show how it can alert the user
to the conflicting information in the summaries. 
The summary statisitcs $\bar{y}$ and $s^2$ are approximately linearly related to each other,
since they both estimate $\eta$, and so we use linear regression for imputation using the 
{\tt R} package {\tt mice} \citep{vanbuuren+g11}.  The {\tt mice} function from this package is used
to generate $M=100$ multiple imputations with the Bayesian linear regression 
method and default settings for other tuning parameters.  For the regression likelihood-free inference, we are using 
the quantile regression forests approach \citep{meinshausen06} adapted to the ABC setting by 
\cite{raynal+mprre18} and available in the {\tt R} package \texttt{abcrf}.  For the random forests regression we use $10,000$ simulations
from the prior, and we tune three parameters of the random forests algorithm (`m.try', `min.node.size' and `sample.fraction') using 
the \texttt{tuneRanger} package \citep{tuneRanger}.  Figure \ref{fig:eg1_abcrf} shows the true posterior density, and 
some random forests estimates of it.  
There are three estimates obtained from the fitted random forests regression
with the full set of features $(\bar{y},s^2)$.  
The first is based on the observed summary statistic for $(\bar{y},s^2)$.  The second 
uses the same fitted regression but replaces the observed summary statistics with $(\bar{y},\widehat{s}^2)$, 
where $\widehat{s}^2$ denotes an imputed value for $s^2$ from the observed $\bar{y}$.  
With imputation the estimate shown is an average over $M$ imputed values.  
The third uses $(\widehat{\bar{y}},s^2)$ where $\widehat{\bar{y}}$ 
is an imputed value for $\bar{y}$ based on the observed $s^2$.
Also shown are density estimates where the regression is fitted using only features $\bar{y}$, and
where the regression is fitted only using features $s^2$.   

\begin{figure}[h]
	\centerline{\includegraphics[width=80mm]{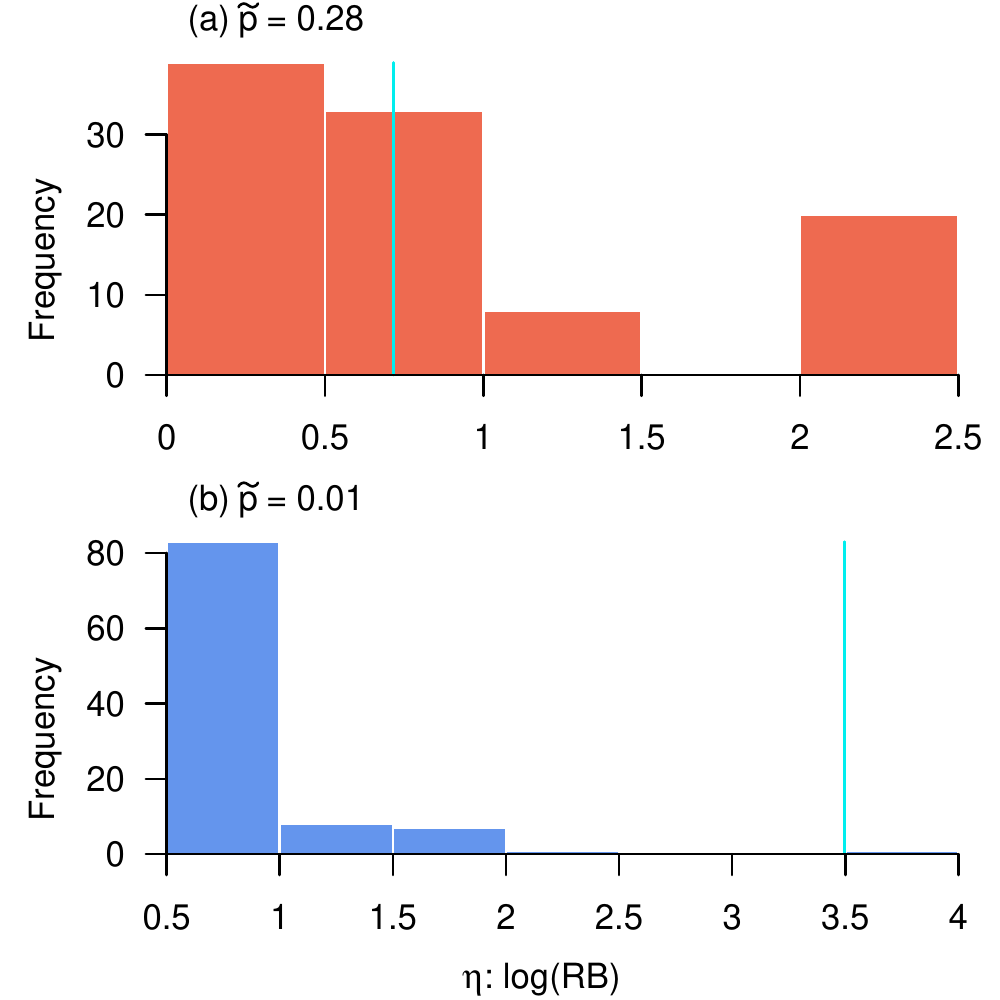}}
	\caption{Observed maximum log relative belief statistic (vertical lines) and histogram of reference distribution values for imputations when $s^2$ is imputed from $\bar{y}$ (top panel) and $\bar{y}$ is imputed from $s^2$ (bottom panel) in Example 5.1. }
	\label{fig:eg1_Poisson_RB}
\end{figure}

We make four observations.  First, using the full summary statistic $(\bar{y},s^2)$ with quantile
regression forests results in an accurate estimate of the true posterior density, in contrast to the ABC
algorithm described in the introduction.  
The reason for this is the variable selection capability of the quantile regression forest approach, 
which results in the summary statistic $s^2$ being ignored as it is not informative given the minimal sufficient
statistic $\bar{y}$.  This is what is expected for our diagnostic, following the discussion of Section 4.3.
Our second observation is that imputing $s^2$ from $\bar{y}$ results in very little
change to the estimated posterior density.  Again this is because the random forest regression fit ignores
$s^2$, and so replacing its observed value with a very different imputation is unimportant.
Our third observation is that imputing $\bar{y}$ from $s^2$ does result in a posterior density with very different
inferential implications. 
Finally, fitting the quantile regression forest with features $(\bar{y},s^2)$ and
then deleting and imputing one of the features for the observed data gives a similar posterior density estimate
to that obtained by fitting the quantile regression with only the feature that was not deleted.  
This is desirable, because if we examine deletion and imputation for many subsets of features 
it may be computationally intensive to refit regression models repeatedly, but the imputation process
is less computationally burdensome.  
Finally, Figure \ref{fig:eg1_Poisson_RB} illustrates our suggested calibration of $R_\infty(S_{B,\text{obs}}|S_{A,\text{obs}})$
for the choices $S_A=\bar{y}$, $S_B=s^2$ and $S_A=s^2$, $S_B=\bar{y}$ and $M^*=100$.  
The vertical lines in the graphs are the observed 
statistic values, and the histograms shows reference distribution values for the statistics for imputations of $S_B$ from $S_{A,\text{obs}}$.  
The estimated tail probability
(\ref{p-value}) corresponds to the proportion of reference distribution values above the observed value.  The change in the posterior density
when $\bar{y}$ is imputed from $s^2$ is surprisingly large (bottom panel).

What would we do in this and similar examples once a conflict is found?  In this example it is natural to
change the model so that the mean and variance need not be the same (using a negative binomial model rather than Poisson, for instance).
Alternatively, if the problem is such that 
the current model is ``good enough'' in the sense that the fitted model reproducing aspects of the data
we care about in the application at hand, we might change the summary statistics so that the conflict is removed while 
retaining the important information.  

\subsection{Stereological extremes}

The next example was discussed in \cite{bortot+cs07} and examines an ellpitical model for diameters of inclusions
in a block of steel.  The size of the largest inclusion is thought to be an important quantity related to the steel strength.
The ellpitical model, which has an intractable likelihood, extends an earlier spherical model
considered in \cite{anderson+c02}.    We do not give full details
of the model but refer to \cite{bortot+cs07} for a more detailed description.

There are 3 parameters in the model.  The first is the intensity $\lambda$ of the homogeneous Poisson point process
of the inclusion locations.  There are also two parameters in a model for the inclusion diameters.  This model is a 
generalized Pareto distribution with scale parameter $\sigma$
and shape $\xi$, for diameters larger than a threshold which is taken as $5\mu$m.  
Writing $\theta=(\lambda,\sigma,\xi)^\top$, the prior density for
$\theta$ is uniform on $[2,200]\times [0,10]\times [-5,5]$.   The observed data consists of $N=112$ 
inclusion diameters observed from a single planar slice.
The summary statistic used consists of $N$ and the quantiles of diameters at
levels $q=0,0.05,0.1,0.2,0.8,0.9,0.95,1$.  Here there may be some interest in whether the simple
Pareto model can fit well for the observations in the lower and upper tail simultaneously.  Quantile-based summary
statistics for ABC were also considered in \cite{erhardt+s16}.  It is expected that the Pareto model will
fit the upper quantiles well, but using only the most extreme quantiles might result in a loss of information.  
Capturing the behaviour of the diameters in the upper tail is most important for the application here, given the relationship
between the largest inclusions and steel strength.

\begin{figure}[h]
	\centerline{\includegraphics[width=\linewidth]{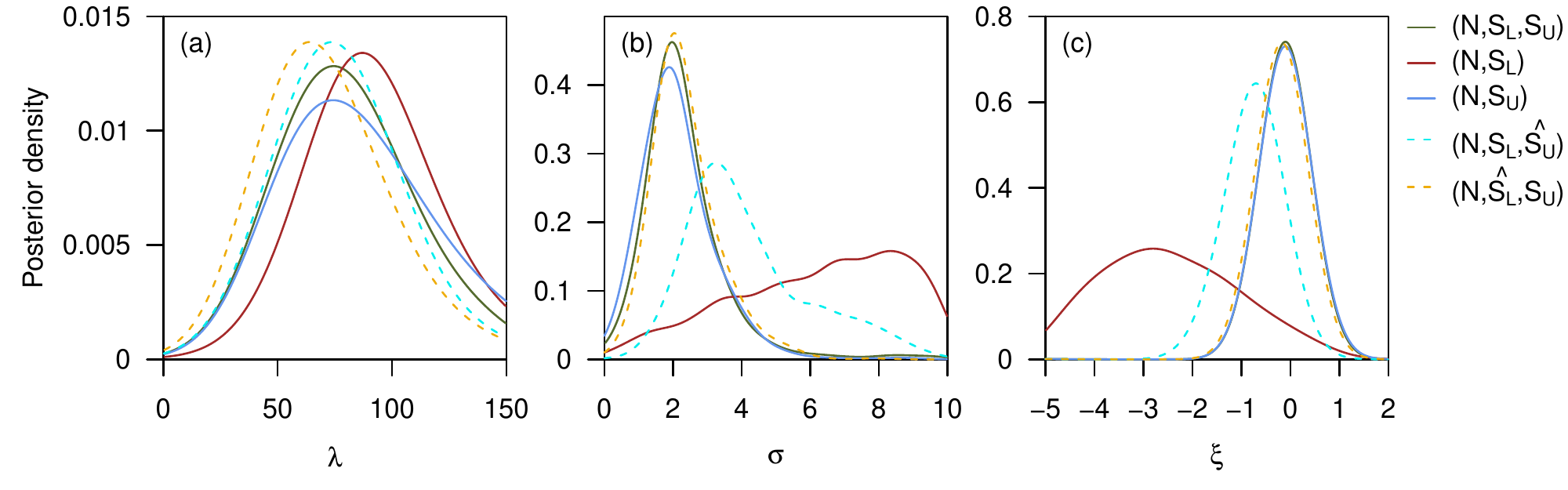}}
	\caption{Estimated marginal posterior densities obtained by quantile regression forests using different summary statistics and imputations of subsets of the summary statistics in Example 5.2.}
	\label{fig:example3}
\end{figure}

We apply our diagnostic method in the following way.  First, we used quantile regression forests to
estimate the marginal posterior densities for each parameter $\lambda$, $\sigma$, $\xi$ separately, and using
the full set of summary statistics, which we denote by $S$. We can write
$S=(N,S_L^\top,S_U^\top)^\top$, where $S_L$ denotes the vector of lower quantiles at levels $q=0,0.05,0.1,0.2$ and
$S_U$ denotes the vector of upper quantiles at levels $q=0.8,0.9,0.95,1$.  In fitting the random forests
regressions we used $500,000$ simulations of parameters and summary statistics from the prior, and 
the random forest tuning parameters were chosen using the \texttt{tuneRanger} package in R.  
We apply our diagnostic by deleting and imputing $S_L$ and $S_U$ respectively to see if there is 
conflicting information in the lower and upper quantiles about the parameters.  
The imputation is done using the \texttt{R} package \texttt{missRanger}, which uses a random
forests approach \citep{mayer19}.  We use the \texttt{missRanger} function in this package with $M=100$ and 
20 trees with other tuning parameters set at the default values.
Figure \ref{fig:example3} shows how the 
estimated marginal posterior densities change for the various parameters upon deletion and imputation
of $S_L$ and $S_U$.  

We make two observations.  First, when the upper quantiles are imputed from the lower ones, the estimated
marginal posterior densities change substantially for $\sigma$ and $\xi$, the parameters
in the inclusion model.  This does suggest that the Pareto model is not able to simultaneously fit the lower
and upper quantiles well.  Given the importance in this application of capturing the behaviour
of the largest diameters, we might remove the lower quantiles from the vector of summary
statistics to obtain a model that is fit for purpose.  Secondly, and unlike our first example, 
the results of deletion and imputation do not match very well with the
estimated posterior densities obtained by refitting the regression for parameter subsets directly.  
However, the estimates based on deletion and imputation are sufficiently good for the diagnostic purpose of 
revealing the conflict, and the computational burden of imputation is much less 
than refitting, given that our regression training set
contains $500,000$ observations, and that algorithm parameters need to be tuned by cross-validation.  

\begin{figure}[h]
	\centerline{\includegraphics[width=\linewidth]{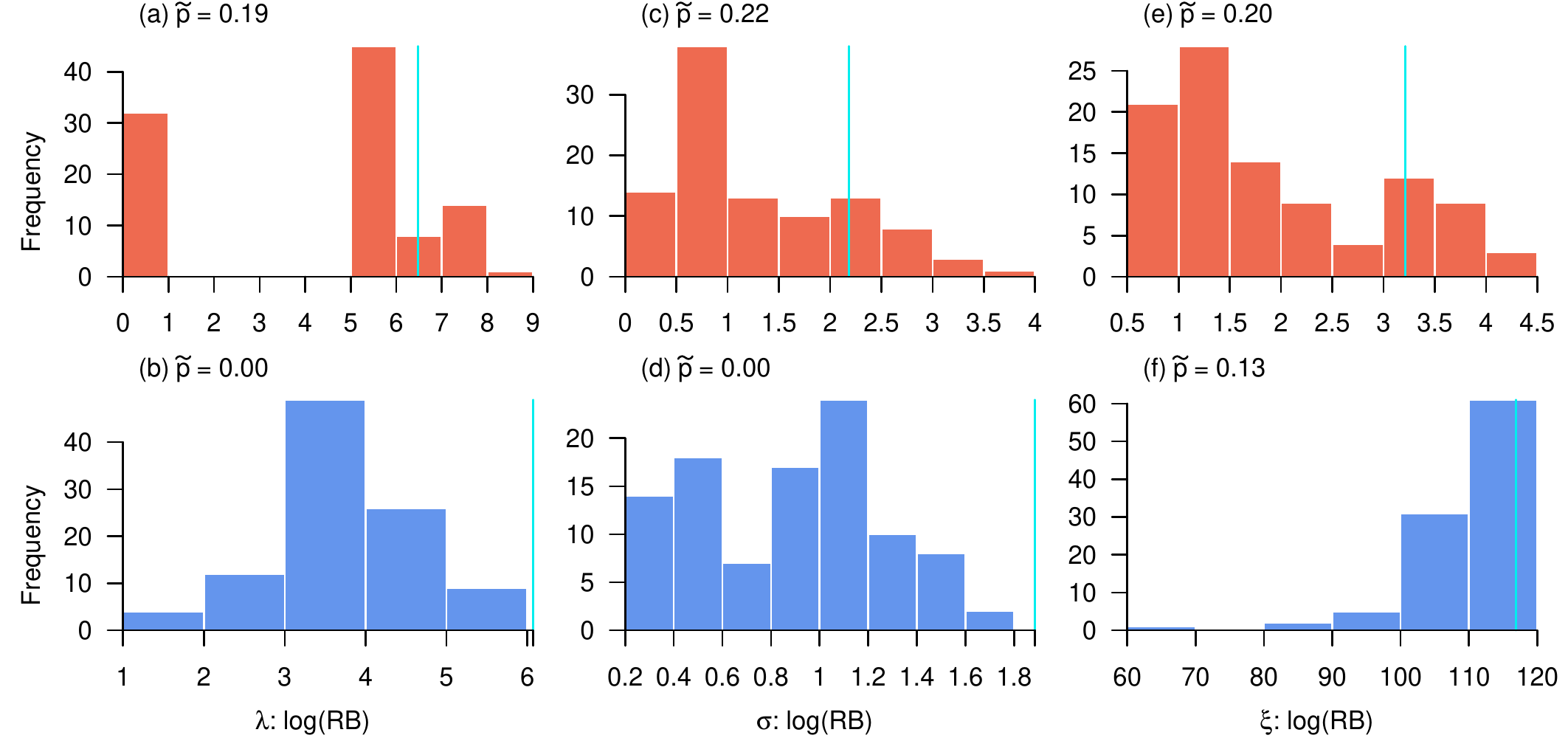}}
	\caption{Observed maximum log relative belief statistic (vertical lines) and histogram of reference distribution values for imputations when $S_L$ is imputed from $(N,S_U)$ (top row) and $S_U$ is imputed from $(N,S_L)$ (bottom row) in Example 5.2.  Different columns correspond to the parameters $\lambda$, $\sigma$, $\xi$.
	\label{fig:example3_EllipticRB}}
\end{figure}

Figure \ref{fig:example3_EllipticRB} shows our suggested calibration of $R_\infty(S_{B,\text{obs}}|S_{A,\text{obs}})$
for the choices $S_A=(N,S_L)$, $S_B=S_U$ and $S_A=(N,S_U)$, $S_B=S_L$ and $M^*=100$.  
The vertical lines in the graphs are the observed 
statistic values, and the histograms shows reference distribution values for the statistics for imputations of $S_B$ from $S_{A,\text{obs}}$.  When $S_U$ is imputed from $S_L$ (bottom row), the change in the estimated posterior density is surprising for
$\lambda$ and $\sigma$.  

\subsection{Ricker model with threshold-autoregressive auxiliary model}

Here we consider a Ricker model \citep{ricker54}, a simple model for the dynamics of animal population sizes 
in ecology.  Likelihood-free inference will be considered using summary statistics which are
point estimates from an auxiliary model with tractable likelihood.  

The auxiliary model is a two-component self-exiciting threshold autoregression (SETAR) model
\citep{tong11}, in which the two auto-regressive components
describe dynamic behaviour of the process at high and low levels respectively.   Maximum likelihood 
estimates of the SETAR model parameters provide the summary statistics.  The use
of an auxiliary models with tractable likelihood is a common way to obtain summary statistics 
for likelihood-free inference - see \cite{drovandi+pl15} for a unified discussion of such approaches.  
We focus on whether there is conflict between the summary statistics for the two autoregressive
components in our auxiliary model.  Our checks can be insightful about the ways in which the Ricker model fails to fit 
well at high and low levels.

Let $N_t$, $t\geq 0$ be a series of unobserved population sizes, and suppose we observe values
$d_t\sim \mbox{Poisson}(\phi N_t)$, where $\phi$ is a sampling parameter.  The series $N_t$ has some
initial value $N_0$ and one-step conditional distributions are defined by
\begin{align*}  
 N_{t+1} & = r N_t\exp(-N_t+e_{t+1}),
\end{align*}
where $r$ is a growth parameter and $e_t\sim N(0,\sigma^2)$ is an independent environmental noise series.  
We write $\theta=(\log \phi,\log r,\log \sigma)^\top$ for the parameters, and 
the prior density for $\theta$ is uniform on the range 
$[11,13]\times [-0.02,0.04]\times [-2,-0.5]$. The prior was chosen 
to achieve a reasonable scale and variation based on prior predictive simulations.  
The Ricker model can exhibit chaotic behaviour when the environmental noise variance is small.  In this case, 
it can be challenging to fit the model using methods which perform state estimation to approximate 
the likelihood.  \cite{wood10} and 
\cite{fasiolo+pw16} discuss further the motivations for using likelihood-free
methods for time series models with chaotic behaviour.  
Model misspecification could be another reason for conditioning
only on data summaries that we care about -- the model for the full data may be considered as a device
for implicitly defining a model for summary statistics, and it may be that we only care about good
model specification for certain summary statistics important in the application.

For a time series $X_t$, $t=0,1,\dots T$, the SETAR model used to obtain summary statistics takes the form
\begin{eqnarray*}
  X_t & = \left\{ \begin{array}{ll}
    a_0+a_1X_{t-1}+a_2X_{t-2}+\epsilon_t,\;\;\;\epsilon_t\sim N(0,\rho^2) & \mbox{if }X_{t-1}<c \\
    b_0+b_1X_{t-1}+b_2X_{t-2}+\epsilon_t,\;\;\;\epsilon_t\sim N(0,\zeta^2) & \mbox{if }X_{t-1}\geq c
    \end{array}\right. .
\end{eqnarray*}
Independence is assumed at different times for the noise sequence $\epsilon_t$.  The parameter
$c$ is a threshold parameter, and the dynamics of the process switches between two autoregressive components
of order 2 depending on whether the threshold is exceeded.  To obtain our summary statistics, we
fit this model to the observed data, and fix $c$ based on the observed data fit.  With this fixed $c$ the SETAR
model is then fitted to any simulated data series $d$ to obtain maximum likelihood 
estimates of the SETAR model parameters, which are the summary statistics denoted by $S=S(d)$.  We write 
$S=(S_L^\top,S_U^\top)^\top$, where
$S_L=(\widehat{a}_0,\widehat{a}_1,\widehat{a}_2,\widehat{\rho})^\top$ and $S_U=(\widehat{b}_0,\widehat{b}_1,\widehat{b}_2,\widehat{\zeta})^\top$ are maximum likelihood estimates
for the autoregressive component parameters for low and high levels respectively.  
The SETAR models are fitted using the \texttt{TAR} package \citep{nieto17} in {\tt R}.
In simulating data from the model there were some cases where there were no values of the series above
the threshold $c$.  Since the number of such cases was small, we simply discarded these simulations.  

The observed data are a series of blowfly counts described in \cite{nicholson54}.  Nicholson's experiments 
examined the effect of different feeding restrictions on blowfly population dynamics.  We use the series for
the adult food limitation condition shown in Figure 3 of \cite{nicholson54}. 
The Ricker model fits the blowfly data poorly, but the example is instructive for illustrating methodology for
model checking.  A much better model is described in \cite{wood10}, based on a related
model considered in \cite{gurney+bn80}.  In the model of \cite{wood10}, the population size is 
a sum of a recruitment and survival process, with the recruitment process related to previous population size
at some time lag.  This model is able to reproduce the strong periodicity and other complex features of 
the data, which the simple Ricker model does not do.

\begin{figure}[h]
		\centerline{\includegraphics[width=\linewidth]{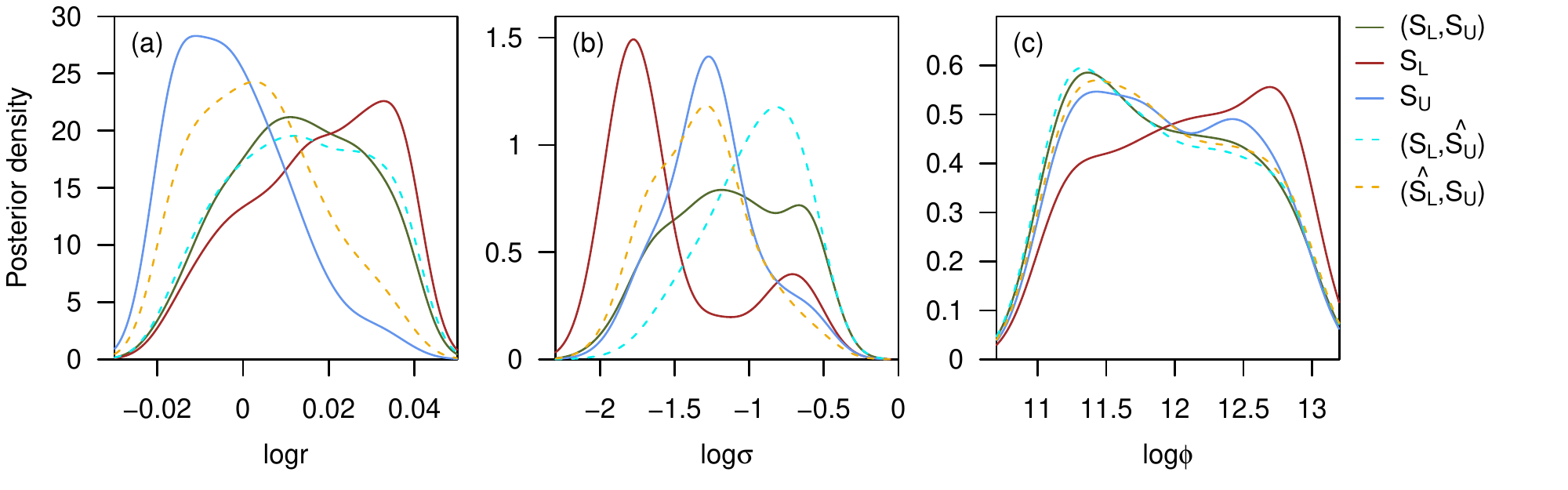}}
		\caption{Estimated marginal posterior densities obtained by quantile regression forests using different summary statistics and imputations of subsets of the summary statistics in Example 5.3.}
		\label{fig:example2_abcrf}
	\end{figure}

We compute ABC posterior estimates for each parameter 
using quantile regression forests, once again tuning
algorithm parameters by cross-validation using the \texttt{tuneRanger}  package.  The quantile regression forests 
are fitted based on $50,000$ values from the prior with corresponding TAR summary statistics.  
We consider fitting the random forests regression using summaries $S$, $S_L$ only and $S_U$ only.  
For the fit with summary statistics $S$ we consider the posterior density estimated using the observed $S$, as
well as values of $S$ where $S_L$ is imputed from $S_U$, and where $S_U$ is imputed from $S_L$.  
The imputation is done using the \texttt{R} package \texttt{missRanger} \citep{mayer19} using the \texttt{missRanger} 
function with $M=100$ and 20 trees with other tuning parameters set at the default values.
The results are shown in Figure \ref{fig:example2_abcrf}.  

We make two observations.  First, the posterior density for $\log R$ is very different
when only $S_U$ is used compared to either $S_L$ only or $S$.  This suggests that the dynamics at high and low levels
of the series are inconsistent with a single fixed value of the growth parameter $R$.
Second, the posterior densities estimated using quantile regression forests
with features $S_U$ only, are estimated quite well by the corresponding estimates using features $S$ and deletion and imputation.
For the posterior densities estimated using $S_L$ only, they are also estimated quite well for $R$, but not the other parameters, using features $S$ and deletion and imputation of $S_U$.  

\begin{figure}[h]
		\centerline{\includegraphics[width=\linewidth]{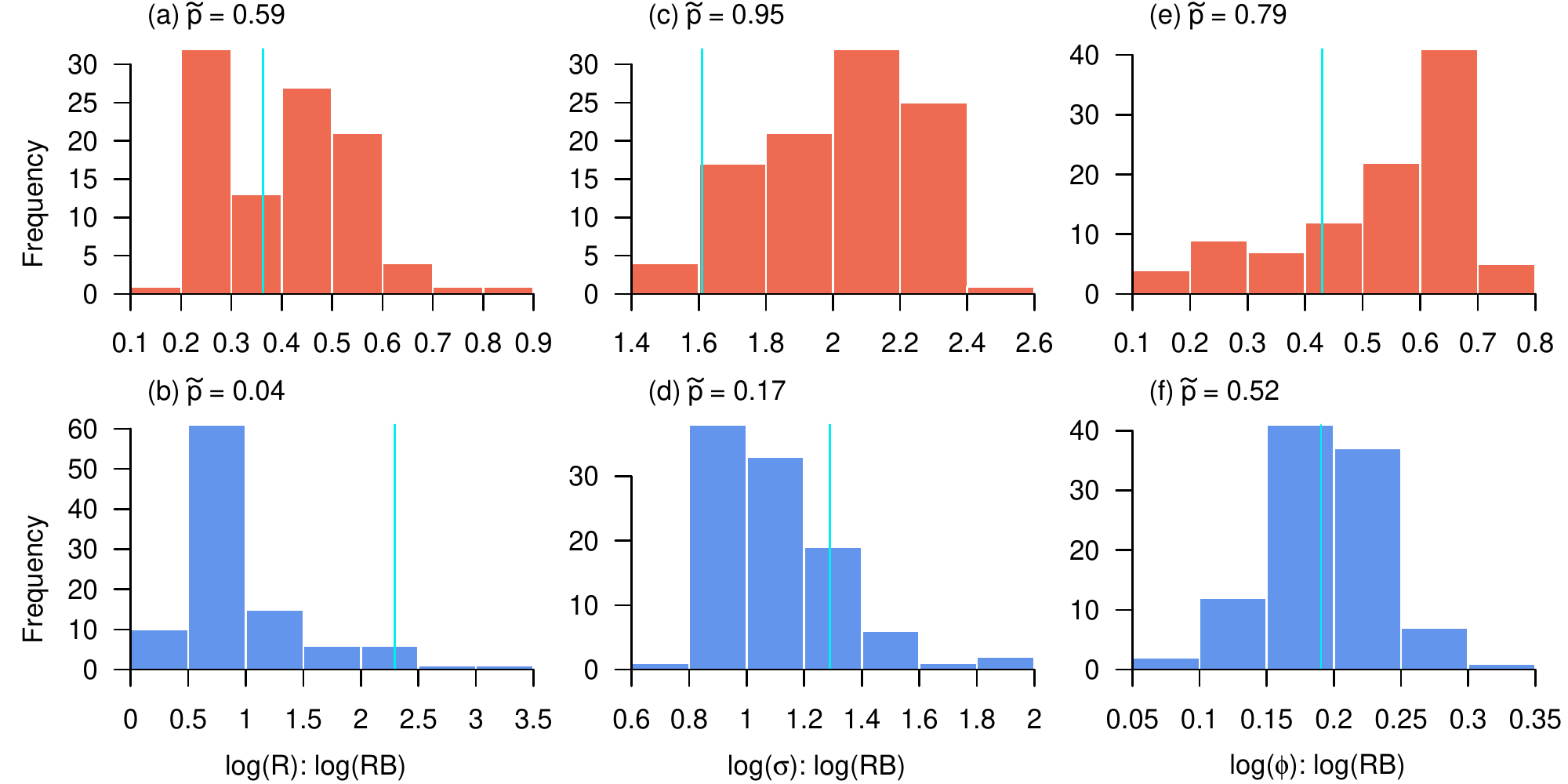}}
		\caption{Observed maximum log relative belief statistic (vertical lines) and histogram of reference distribution values for imputations when $S_L$ is imputed from $S_U$ (top row) and $S_U$ is imputed from $S_L$ (bottom row) in Example 5.3.  Different columns correspond to the parameters $R$, $\sigma$, $\phi$.}
		\label{fig:example2:plotRickerRB}
	\end{figure}

Figure \ref{fig:example2:plotRickerRB} shows our suggested calibration of $R_\infty(S_{B,\text{obs}}|S_{A,\text{obs}})$
for the choices $S_A=S_L$, $S_B=S_U$ and $S_A=S_U$, $S_B=S_L$.  The vertical lines in the graphs are the observed 
statistic values, and the histograms shows reference distribution values for the statistics for imputations of $S_B$ from $S_{A,\text{obs}}$.  When $S_U$ is imputed from $S_L$ (bottom row), the change in the estimated posterior density
for $\log R$  is surprising.  
	
\subsection{Ricker model without summary statistics:  convolutional neural networks}

We now consider using the whole time series as the feature vector in the Ricker model, and use a convolutional neural network
as the regression method.  Convolutional networks are not described in detail here; the reader is referred
to \cite{polson+s17} and \cite{fan2019} for introductory discussions suitable for statisticians.
\cite{dinev+g18} considered the use of convolutional networks for automated summary statistic choice for 
likelihood-free inference for time series and
showed that a single architecture can provide good performance in a range of problems.  
In fitting the convolutional network we use a squared error loss, and after obtaining point estimates
for the network parameters and response variance we use a plug-in normal predictive distribution.

\begin{figure}[h!]
\centerline{\includegraphics[width=70mm]{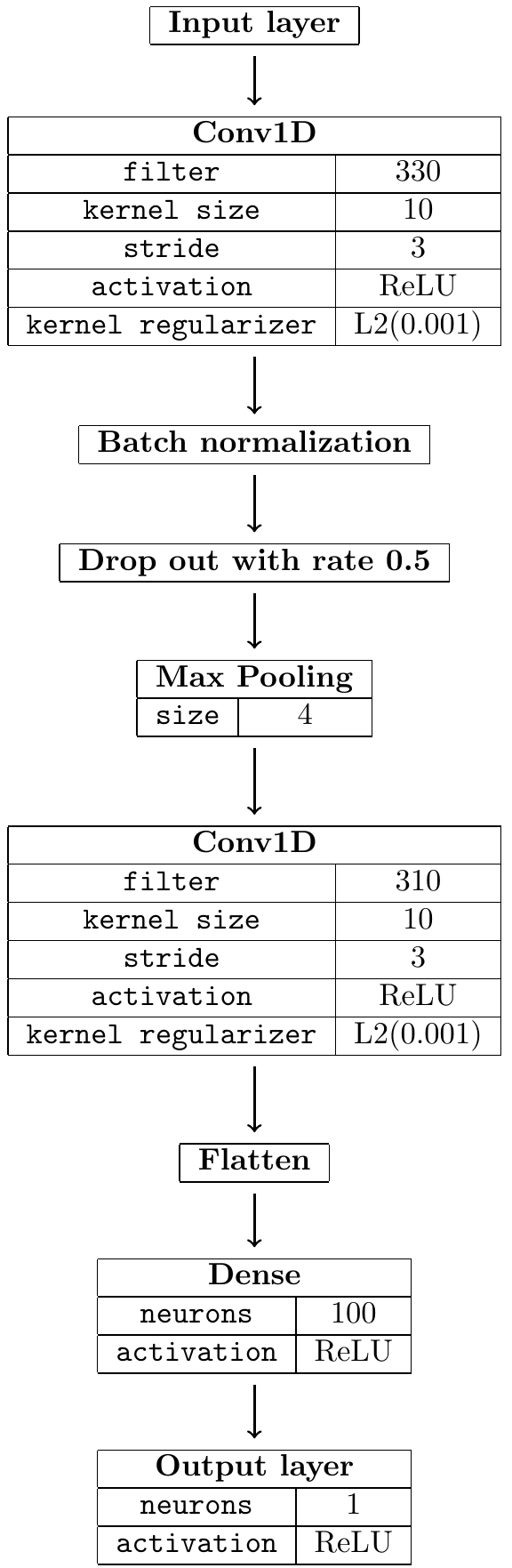}}
\caption{CNN model used for Example 5.4. The architecture used is similar to that
in \cite{dinev+g18}}.
\label{fig:cnn}
\end{figure}

Writing, as before, $\eta$ for both the full parameter $\eta=(\log\phi,\log r, \log \sigma)^\top$, as well as some scalar
function of interest (where the meaning intended is clear from the context), we estimate the marginal posterior density
for $\eta$ from samples $(\eta^{(i)},d^{(i)})\sim p(\eta)p(d|\eta)$, $i=1,\dots, N$.  We used $N=10,000$ below.  
The regression model is
\begin{align*}
  \eta^{(i)}& = f_w(d^{(i)})+\epsilon^{(i)},
\end{align*}
for errors $\epsilon^{(i)}$, where $f_w(\cdot)$ is a convolutional neural network with weights $w$.  
Figure \ref{fig:cnn}
shows the architecture used, which is based on the architecture shown in Figure 6 (a) of \cite{dinev+g18}.  
Training of the network was done using the Keras API for Python, using the Adam optimizer with default settings.  The model was
trained for 50 epochs with a batch size of 64.   
After estimating $(w,\sigma_\epsilon^2)$ as $(\widehat{w},\widehat{\sigma}_\epsilon^2)$, the estimated
posterior density for data $d$ is $N(f_{\widehat{w}}(d_{\text{obs}}),\widehat{\sigma}_\epsilon^2)$.

We wish to develop some diagnostics useful for measuring the influence of the observation at time $t$ on
inference.  A method is developed here similar to that of 
\cite{zintgraf+caw17} for measuring the influence of a pixel in image classification.  
Deletion of ``windows'' of the raw series containing $t$ can be considered (i.e. $S_B$ are
the values of the series in a window of width $k$, say,
containing $t$, and $S_A$ is the remainder of the series) and then we consider averages of the 
statistics $R_\infty (S_B|S_A)$ over all windows $B$ containing $t$.  In this example we will
consider windows of size $k=4$, and window values are imputed from one neighbouring value
to the left and right of the window.  At the boundary points where
only one neighbouring value is available we use this.
The method is described more precisely in the Appendix, as well as our approach to 
the multiple imputation.   When computing the maximum log relative belief statistic 
(\ref{logrbstat}), we do not truncate the normal predictive
densities for the regression model to the support of the parameter space, 
but we do compute the supremum over this support.

\begin{figure}[h]
		\centerline{\includegraphics[width=\linewidth]{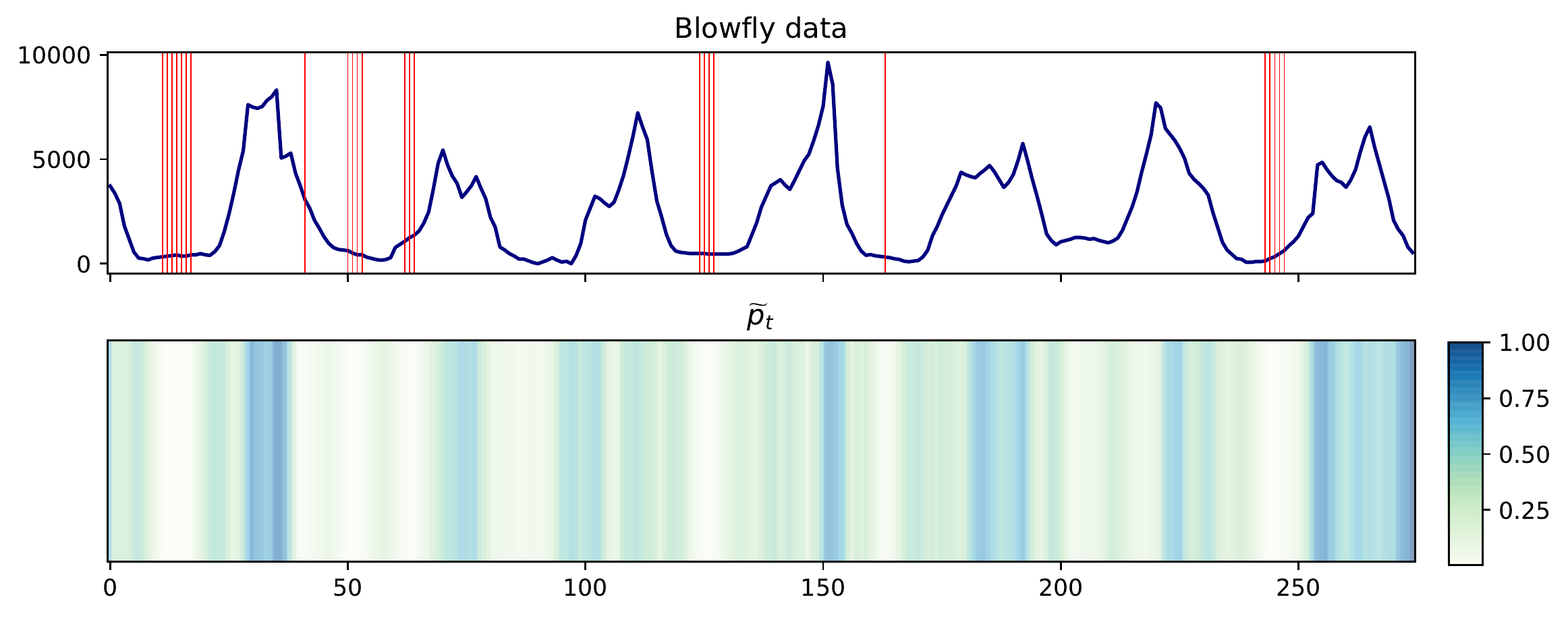}}
		\caption{Observed series (top panel) with times where $\widetilde{p}<0.05$ marked as vertical lines, together with
		plots of $\widetilde{p}$ versus time (bottom panels), for parameter $\log \sigma$ in Example 5.4.}
		\label{convnetresults}
	\end{figure}

For the parameter $\log \sigma$, Figure \ref{convnetresults} shows a 
plot of the raw series values with points $t$ with calibration probability $\widetilde{p}_t$ for our diagnostic less than 
$0.05$ indicated
by vertical lines (top panel), and a plot of the $\widetilde{p}_t$ values themsleves indicated by coloured bands (bottom panel).  
Some of the smallest values in the series around four of the local minima are influential for the estimation of $\log \sigma$.
Similar plots for the other two parameters did not show any points where the calibration probabilities are very small, except
for one point near the boundary for $\log \phi$;  however, in that case the failure of the imputation method to handle 
boundary cases well may be an issue.   
Compared to our previous analysis, the approach using convolutional networks examines misfit for temporally localized
parts of the series.  Poor fitting regions 
seem to particularly affect the environmental noise variance parameter that describes variation unexplained by the model
dynamics.

\section{Discussion}

We have discussed a method for detecting the existence of conficting summary statistics using regression approaches
to likelihood-free inference.  The approach uses a recent idea from the interpretable machine learning literature, based on
deletion and imputation of part of a feature vector to judge the importance of subsets of features for
the prediction.  In likelihood-free regression approximations, where the regression predictive distribution is the 
estimated posterior distribution, the method estimates the posterior distribution conditioning only on a subset of 
features, and compares this with the full posterior distribution.  The deletion and imputation process is less
computationally burdensome than fitting a flexible regression model for different feature subsets directly.  The 
approach can be related to prior-data conflict checking methods examining the consistency 
of different sources of information, and in this interpretation the posterior distribution based on a subset
of the features is considered as the prior, which is updated by the likelihood term for the remaining features.  
Based on this connection, a way to calibrate the diagnostics is suggested.  

A concern expressed in the work of \cite{zintgraf+caw17} was possible sensitivity of their 
procedure to the imputation method used.  
That is a concern in our applications also, although we used a number of different imputation methods in the examples and 
did not find this sensitivity to be a problem for reasonable choices.  
The diagnostics described are computed using regression approaches to likelihood-free inference, but they
may be useful even if another likelihood-free inference approach is employed
for inferential purposes, or in the case where the likelihood is tractable.  
We believe that insightful methods for model criticism are very important, 
since they can result in meaningful model expansions and refinements, 
and ultimately more appropriate inferences and decisions.

\section*{Acknowledgements}

Michael Evans was supported by a grant from the Natural Sciences and Engineering Research Council of Canada. 

\section*{Appendix}

\subsection*{Details of diagnostic for the example of Section 5.4}

We describe how we implement our diagnostic for the example of Section 5.4.  Roughly speaking, all windows $B$ of width $k$
are considered including a time $t$, and then we average $R_\infty (S_B|S_A)$ over 
$B$ where $S_B$ consists of the series values in $B$ and $S_A$ is the remaining values.

To make the method precise we need some further notation.  Let $d=\{d_i:1\leq i\leq T\}$ denote a time
series of length $T$.  For some subset $C$ of the times, $C\subseteq \{1,\dots, T\}$, we write 
$d(C)=\{d_i:i\in C\}$ and $d(-C)=\{d_i:i\notin C\}$.  Let $t\in \{1,\dots, T\}$ be a fixed time.  
Let $W_t^K=\{C_1^{t,k},\dots, C_{n_{t,k}}^{t,k}\}$ denote the set of all windows of width $k$ containing $t$ of
the form $\{l,\dots, l+k-1\}$ for some $l$.  
For each $j=1,\dots, n_{t,k}$ suppose that $d^{t,k}_j$ is some time series of length $T$,
and write $d_.^{t,k}=(d_1^{t,k},\dots, d_{n_{t,k}}^{t,k})$.  
Write $d_{obs}^{t,k}$ for the value of $d_.^{t,k}$ where $d_j^{t,k}=d_{\text{obs}}$ for all $j=1,\dots, n_{t,k}$ where $d_{\text{obs}}$ is the observed series.  Let $d_.^{t,k,*}$ denote the value of $d_.^{t,k}$ where 
$d_j^{t,k,*}(-C_j^{t,k})=d_{obs}(-C_j^{t,k})$ and 
$d_j^{t,k,*}(C_j^{t,k})\sim p(d(C_j^{t,k})|d_{\text{obs}}(-C_j^{t,k}))$ i.e. the observation for $d_j^{t,k,*}$ in
$C_j^{t,k}$ are generated from the conditional prior predictive given the observed value for the remainder of
the series.  The draws $d_j^{t,k,*}(C_j^{t,k})$ are independent for different $j$.  Let
\begin{align*}
  R^{t,k}(d_.^{t,k}) & = \frac{1}{n_{t,k}} \sum_{j=1}^{n_{t,k}} R_\infty(d_j^{t,k}(C_j^{t,k})|d_j^{t,k}(-C_j^{t,k})). 
\end{align*}
We base our diagnostic on $R^{t,k}(d_{\text{obs}}^{t,k})$, calibrated by
\begin{align}
  p_t & = P(R^{t,k}(d_.^{t,k,*})\geq R^{t,k}(d^{t,k}_{\text{obs}})),  \label{pt-value}
\end{align}
and estimate (\ref{pt-value}) by 
\begin{align}
  \widetilde{p}_t & = \frac{1}{M^*} \sum_{i=1}^{M^*} I(R^{t,k}(d_.^{t,k,i})\geq R^{t,k}(d_{\text{obs}}^{t,k})), \label{est-pt-value}
\end{align}
where $d_.^{t,k,i}$, $i=1,\dots, M^*$ are approximations of draws of $d_.^{t,k,*}$ based on imputation, 
i.e. we have imputed $d_j^{t,k,*}(C_j^{t,k})$ from $d_{\text{obs}}(-C_j^{t,k})$ independently for each $j$ and 
$i$.  

\subsection*{Details of the imputation method for the example of Section 5.4}

\begin{figure}[h!]
\centering
\includegraphics[width=0.8\textwidth]{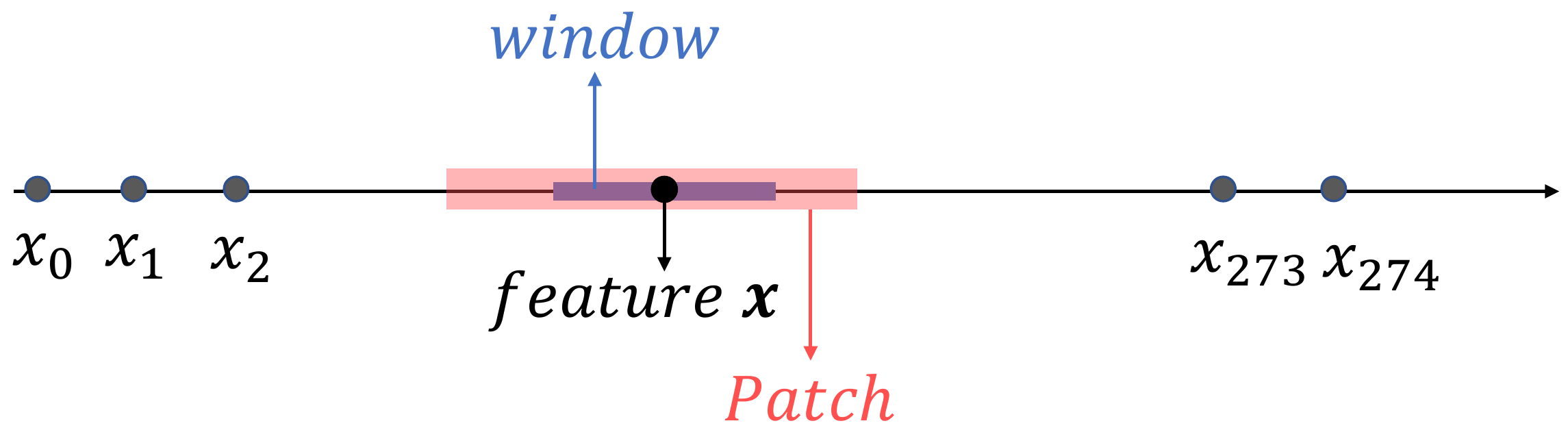}
\caption{Imputation and conditional sampling of features for Example 5.4. Blue-shaded area is the window of length $k$ which we impute in the series. Red-shaded area is the larger window to which a multivariate normal model is fitted based on the training set observations. Normal noise is added to a conditional mean imputation using the covariance matrix for the observations in the blue window, conditional on remaining observations in the red patch.}
\label{fig:windowpatch1d}
\end{figure}

Figure \ref{fig:windowpatch1d} illustrates the idea behind
the window-based imputation we use in the example of Section 5.4.  
We need to impute values of the series for a window of size $k$ which has been deleted (indicated by the blue region in
the figure).
A larger window around the one of width $k$ is considered (red patch in the figure).  
To impute, we first obtain a mean imputation using the functions in the \texttt{imputeTS} package \citep{steffen+b17}.  
In particular, for imputing a conditional mean we use the {\tt na_interpolation} function in \texttt{imputeTS} with 
spline interpolation and default settings for other tuning parameters.  For multiple imputation, we add noise
to the conditional mean by fitting a stationary Gaussian autoregressive model of order one to the observed series
and then consider zero mean Gaussian 
noise, where the covariance matrix of the noise is the conditional covariance matrix of the autoregressive process
in the blue region given the remaining observations in the red patch.   Note that although the series values are
counts, these counts are generally large and we treat them as continuous quantities in the imputation procedure.

\bibliographystyle{chicago}
\bibliography{references}

\end{document}